\documentclass[prl,twocolumn,nofootinbib,longbibliography]{revtex4-1}
\usepackage{graphicx}
\usepackage{color}
\usepackage{epsfig}
\usepackage[caption=false]{subfig}
\usepackage{pdfpages}

\begin{document}
\title{Pinning Susceptibility : A Novel Method to Study Growth of Amorphous 
Order in Glass-forming Liquids}
\author{Rajsekhar Das$^1$}\thanks{These authors contributed equally.}
\author{Saurish Chakrabarty$^2$}\thanks{These authors contributed equally.}
\author{Smarajit Karmakar$^1$}
\email{smarajit@tifrh.res.in}
\affiliation{$^1$TIFR Center for Interdisciplinary Science, 
Tata Institute of Fundamental Research, Narsingi, Hyderabad 500075,
India,\\
$^2$International Centre for 
Theoretical Sciences, Tata Institute of Fundamental Research, 
Shivakote, Bangalore, 560089, India}

\begin{abstract}
Existence and growth of amorphous order in supercooled liquids approaching 
glass transition is a subject of intense research. Even after decades of 
work, there is still no clear consensus on the molecular mechanisms 
that lead to a rapid slowing down of liquid dynamics approaching this putative 
transition. The existence of a correlation length associated with amorphous 
order has recently been postulated and also been estimated using multi-point 
correlation functions which cannot be calculated easily in experiments. 
Thus the study of growing amorphous order remains mostly restricted to 
systems like colloidal glasses and simulations of model glass-forming liquids.
In this Letter, we propose an experimentally realizable yet simple correlation 
function to study the growth of amorphous order. We then demonstrate the 
validity of this approach for a few well-studied model supercooled liquids 
and obtain results which are consistent with other conventional methods.
\end{abstract}
\maketitle

Glasses are ubiquitous in nature and are of immense practical importance in 
our day-to-day lives as well as in modern technology. In spite of knowing their 
presence and usefulness from the early history of mankind, the nature of the 
glassy state and the glass transition, where the viscosity of a liquid increases 
by many orders of magnitude within a narrow temperature or density window, 
still puzzles the scientific community and is believed to be one of the 
major unsolved problems in condensed matter physics \cite{09Cav, 11BB, 
14KDS, 00Ediger, 05Berthier, 06BBMR, 08BBCGV, 09KDS, 15KDSRoPP}. 
The viscosity of a glass forming liquid increases so dramatically that
one is often tempted to believe that viscosity probably diverges at
a certain critical temperature associated with a thermodynamic phase
transition. Thus, not surprisingly, there are two types of theories on
glass transition. 
The first type assumes that the (ideal) glass transition is a
thermodynamic phase transition and glassy states are believed 
to be a thermodynamic state of matter. This approach is taken by the 
Random First Order Transition 
(RFOT) theory~\cite{RFOT,RFOT1,RFOT2}. The other approach is to
consider the glass transition to be a purely dynamic phenomenon
and the slowdown of dynamics is thought to be a result of an
ever increasing number of self-generated kinetic constraints with 
supercooling\cite{kcm,11keysPRX,shreyas2014pinning}. 

In spite of all the differences in opinion about the existence of a true 
thermodynamic glass transition, there is a consensus about the 
existence and growth of correlation lengths along with the rapid 
increase in viscosity or relaxation time\cite{00Ediger,05Berthier, 
08BBCGV,09KDS,11BB,14KDS,15KDSRoPP}. Recently there have been a lot of 
progress in identifying at least two different length scales -- {\bf
(a)} a dynamic length scale akin to the length scale characterizing the 
heterogeneity present in the dynamics of supercooled liquids 
\cite{00Ediger,05Berthier,09KDS} and {\bf (b)} a static
length scale of the so called ``amorphous order''  
\cite{08BBCGV, 12KLP,DelGadoOttiner,12HMR,GKPP15,kurchanLevine, 13BKP}. 
Simple two-point density correlation functions have been shown to be
unsuitable for the identification of the build up of these
correlations in the supercooled liquids. A four point susceptibility,
$\chi_4(t)$ and its corresponding structure factor, $S_4(q,t)$ (see
\cite{SI} for definitions) \cite{cdggrowing,02lacevic,09KDS} were
found to be sensitive to these correlations in glassy liquids. The
intricate nature of the correlation functions limit the applicability
of them only to systems where the dynamical details of individual constituent molecules 
are available, $e.g.$ colloidal glasses \cite{ref:weeksscience,dhexp2} and 
{\em in silico} model glass-forming liquids \cite{dhnum,14KDS}. 
The point-to-set correlation function\cite{08BBCGV,12HMR}, finite size 
scaling of stiffness of local potential energy landscape 
\cite{12KLP,GKPP15,13BKP} and other measures of 
local order \cite{kurchanLevine,RoyallWilliamsPhysRep2015} played an 
important role in identifying the length scale associated with the 
amorphous order. These methods also require microscopic details of the
system in order to calculate the correlation length and are thus not
useful for experiments of a variety of molecular glass forming liquids.

Some recent proposals to study the growth of amorphous
order using higher order non-linear susceptibilities have yielded encouraging 
results. In Refs. ~\cite{nonLinearSusPRL2010, nonLinearSusPRL2012, 
nonLinearSus}, higher order non-linear dielectric susceptibilities
were measured for a few molecular glasses at different temperatures
and it was shown that the third and fifth order non-linear dielectric
constants were sensitive to the growth of amorphous order. It is
important to note that measuring the higher order non-linear
dielectric susceptibilities is extremely difficult as they are many
orders of magnitude smaller than the leading linear
contribution. Special experimental techniques were needed to estimate
these higher order susceptibilities
\cite{nonLinearSusPRL2010, nonLinearSusPRL2012}.
Since $\chi_4(t)$ cannot be estimated directly in experiments, an indirect approach
is taken. Its peak value is estimated using an approximate equality
$\chi_4^P \simeq k_B T^2(\chi_T^P)^2/c_P$,  
where $\chi_T$ ($\equiv \partial Q/\partial T$), the temperature, $T$ being
the control variable
and $Q$ the two-point density overlap correlation function to be
  defined later.
$k_B$ is the Boltzmann constant and $c_P$ is the
specific heat of the system. In Ref. ~\cite{05Berthier}, frequency dependent
$\chi_T$ was measured for pure glycerol in a desiccated Argon environment
in order to eliminate moisture absorption, using capacitive dielectric 
spectroscopy close to the calorimetric glass transition temperature. 
Similarly $\chi_\phi \equiv \partial Q/\partial \phi$, where $\phi$ is the 
packing fraction is measured using dynamic light scattering (DLS) 
experiments on colloidal glass formers for which the packing fraction is 
the relevant control parameter. It was shown that the dynamic length scale 
does grow as the glass transition is approached
and  becomes (assuming exponent $\eta$ to be in the range $2$ to $4$) 
$\sim 1.5nm$ at the lowest temperature studied ($T=192K$) for glycerol. 
In short, all these methods are experimentally delicate and indirect 
when it comes to estimating the growth of cooperativity in supercooled 
liquids.

In this Letter, we propose a very simple correlation function which 
directly estimates the growth of amorphous order and the associated static
length scale without involving unknown fitting parameters. The idea came
from the recent interest in dynamics of supercooled liquids and
glass transition with quenched random pinning sites. It was claimed in 
Refs. ~\cite{okim2015,cammarotaPinning,13KB} that an ideal glass state
can be obtained by introducing quenched disorder in the form of pinned
particles in supercooled liquids. Although the existence of such an
ideal glass state is still debated \cite{CKD15srep,15CKDpnas}, random
pinning has become a diagnostic tool to study the dynamics of
the supercooled liquids and to test the validity of existing theories
of glass transition \cite{cdkd16}.
\begin{figure}
\begin{center}
\hspace{-.1in}\includegraphics[width=0.4\textwidth]{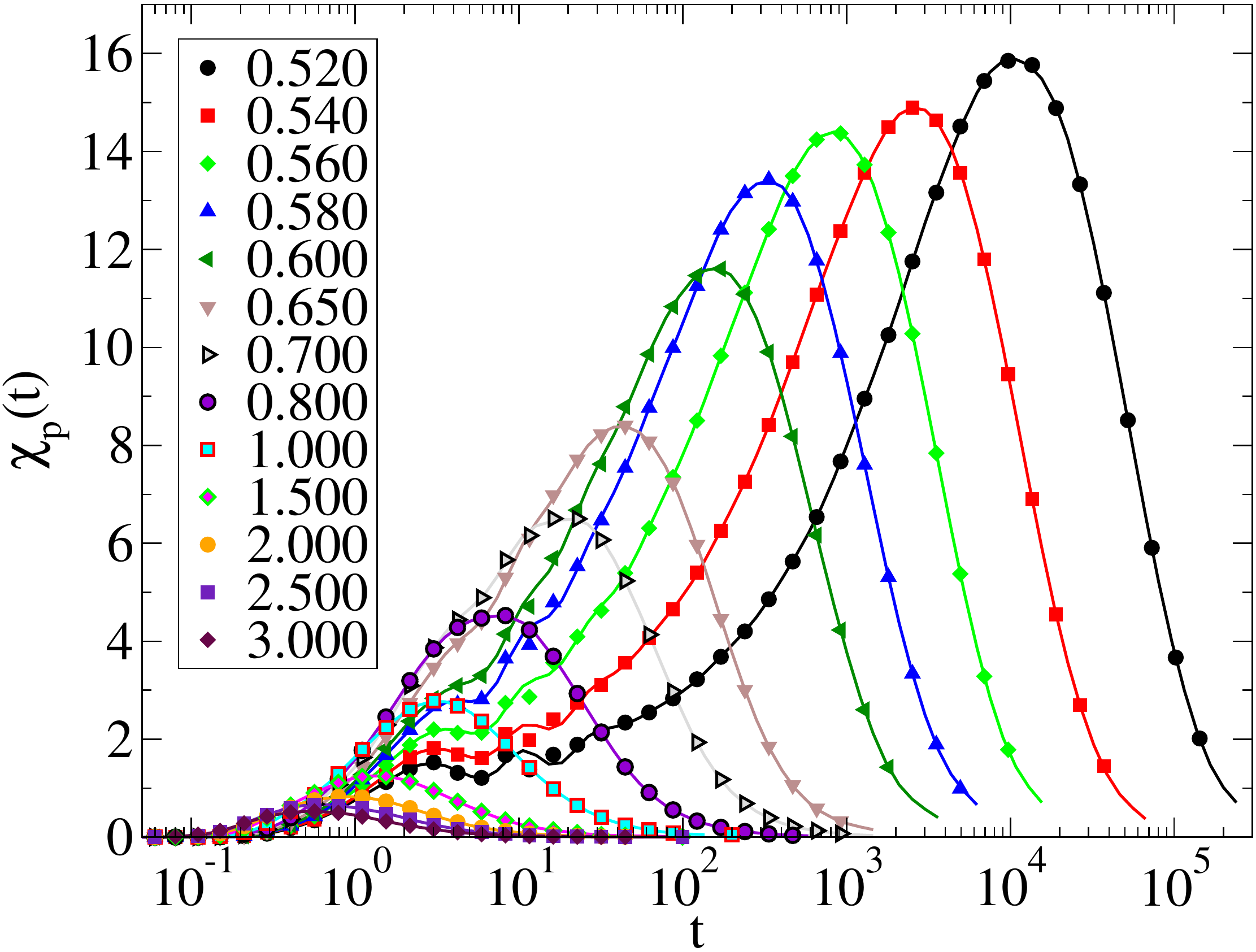}\vskip 0.1cm
\includegraphics[width=0.39\textwidth]{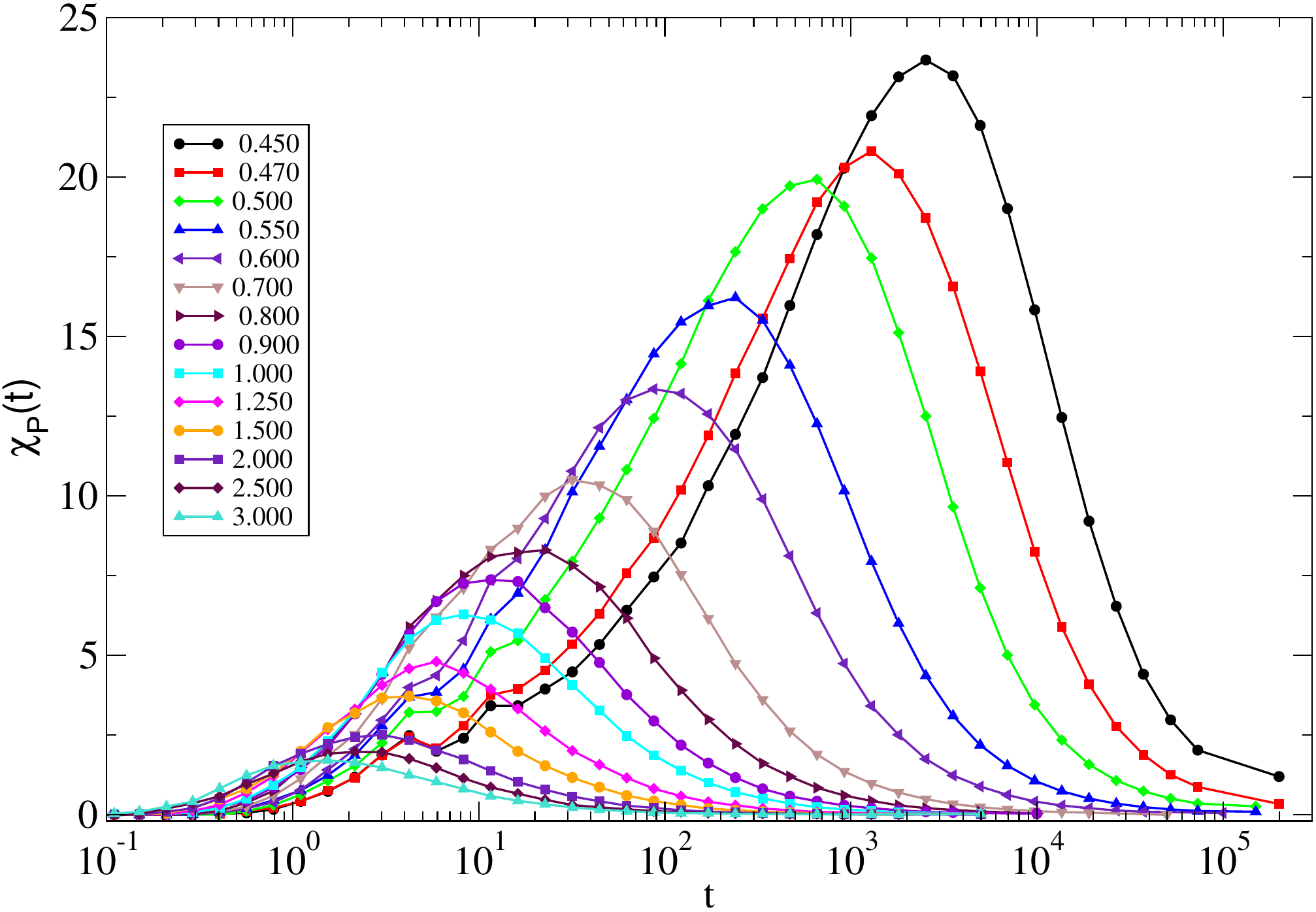}
\caption{{\bf Pinning Susceptibility:} Time-Temperature dependence of 
$\chi_p(t)$ for 3dR10 model (top panel) and 2dR10 model (bottom panel). 
Peak height of $\chi_4(t)$ which directly measures the correlation volume 
of amorphous order, grows monotonically with decreasing temperature.}
\label{chiRho3dKA}
\end{center}
\end{figure}
Similar to $\chi_T$ and $\chi_{\phi}$, in the context of random
pinning, we define the ``pinning susceptibility'' as
\begin{eqnarray}
\chi_p(T,t) = \left.\frac{\partial Q(T,c,t)}{\partial c}\right |_{c=0}
\end{eqnarray}
where $c$ is the fraction of pinned particles and the 
overlap correlation function $Q(T,c,t)$ is defined as,
\begin{equation}
Q(T,c,t) = \frac{1}{(1-c)N}\left[\left<\sum_{i=1}^{(1-c)N}w\left( |\vec{r}_i(t) - \vec{r}_i(0)|\right)\right>\right].
\end{equation}
The window function $w(x)=\Theta(a-x)$, $\Theta(x)$ being the Heaviside 
step function. The square brackets indicate averaging over different 
realizations of the pinned particles and angular brackets stand for ensemble 
average (see Ref. ~\cite{SI} for details). The pinning susceptibility
of a liquid measures the sensitivity of $Q(0,t)$ 
(we drop the temperature argument for brevity)
to small
changes in the number of pinned particles \cite{15sethnaChiPjamming}. 
$\chi_p$ has a peak at a time comparable to 
$\tau_{\alpha}$ [defined as $Q(0,\tau_{\alpha})=1/e$] 
and is zero for both short and long times.

We calculate the pinning susceptibility for 
four different models systems in both two and three dimensions. 
Following existing literature, the systems
in three dimensions are referred to as the ``3dKA'' and ``3dR10'' models 
and those in two dimensions as the ``2dmKA'' and ``2dR10'' models \cite{cdkd16} 
(see Ref. ~\cite{SI} for complete details). In Fig.~\ref{chiRho3dKA}, 
we plot $\chi_p$ for the 3dR10 and 2dR10
models for different temperatures. The function shows features very similar to 
the four-point susceptibility,
$\chi_4(t)$. It has a peak at time scale close to $\tau_{\alpha}$ and 
the peak height grows with decreasing temperature. 
Scaling arguments presented later in the text show that the peak 
height of this newly defined ``pinning susceptibility'', $\chi_p$ is directly 
proportional to $\xi_p^d$, where $\xi_p$ is the static length scale
obtained using random pinning \cite{CKD15srep,15CKDpnas,cdkd16} and
$d$ is the number of spatial dimensions. This pinning length scale, $\xi_p$ is
in turn related to the static length scale, $\xi_s$ associated with the
amorphous order as $\xi_s \sim \xi_p^{d/(d-\theta)}$, where $\theta$ is 
the surface tension exponent in RFOT theory.

The maximum of $\chi_p(t)$, $\chi_p^{max}(T)$ appears at time 
$t \sim \tau_{\alpha}$, so
$\chi_p^{max}(T)\approx \chi_p[T,\tau_\alpha(T)]$.
At times much larger compared to the short time $\beta$-relaxation time 
scale ($i.e.$, comparable to $\tau_\alpha$ time scale), $Q(c,t)$ 
can be approximated very well by stretched exponential functions as,
\begin{eqnarray}
Q(c,t) \sim \exp\left[-\left(\frac{t}{\tau_\alpha(c,T)}\right)^\beta\right],
t\gg\tau_\beta.
\label{stretchedExp}
\end{eqnarray}
This is not very crucial approximation for this analysis 
though
there exist
huge amount of experimental and numerical data to support this 
approximation. $\beta$ is the stretching exponent. 
Differentiating Eq.~\ref{stretchedExp} with respect to $c$ and setting 
$t\sim\tau_\alpha$, we get,
\begin{eqnarray}
\chi_p^{max}(T) \sim 
\left.\frac{\beta}{e\tau_\alpha(c,T)}
\frac{\partial\tau_\alpha(c,T)}{\partial c}\right|_{c=0}
\label{chiRhoTau}
\end{eqnarray}
Assuming that $\chi_p$ attains its maximum value at
$t\sim\tau_\alpha$, the above equation gives us the value of the peak in
$\chi_p(T,t)$ for a given temperature. In Refs.~\cite{CKD15srep,15CKDpnas,cdkd16}
it was shown that for small $c$, the relaxation time
$\tau_{\alpha}(c,T)$ obeys the following scaling relation.
\begin{equation}
\log\left[\frac{\tau_{\alpha}(c,T)}{\tau_{\alpha}(0,T)}\right] = 
f\left[c\xi_p^d(T)\right].
\label{scalingFn}
\end{equation}
Here, the scaling function $f(x)$ must go to zero as the argument goes to 
zero, so in the limit $x\to 0$, $f(x)$ can be approximated as $f(x) \sim x$. The 
validity of the scaling function is demonstrated in Refs.
\cite{CKD15srep,15CKDpnas,cdkd16} and also found to describe experimental 
data very well \cite{colloidalExptPRL}.
Substituting Eq.~\ref{scalingFn} into Eq.~\ref{chiRhoTau}, we get,
\begin{eqnarray}
\chi_p^{max}(T) = \frac{\beta}{e}\xi_p^d(T).
\end{eqnarray}
Note that the pinning susceptibility is directly proportional to the $d^{th}$ 
power of the pinning length scale apart from a weak non-singular dependence 
of the stretching exponent $\beta$ on $T$.
In other words,
\begin{eqnarray}
\xi_p^d(T)\propto\max_t\left[\lim_{c\to0}
\frac{Q(T,c,t)-Q(T,0,t)}{c}\right].
\end{eqnarray}
\begin{figure}
\begin{center}
\includegraphics[width=0.24\textwidth]{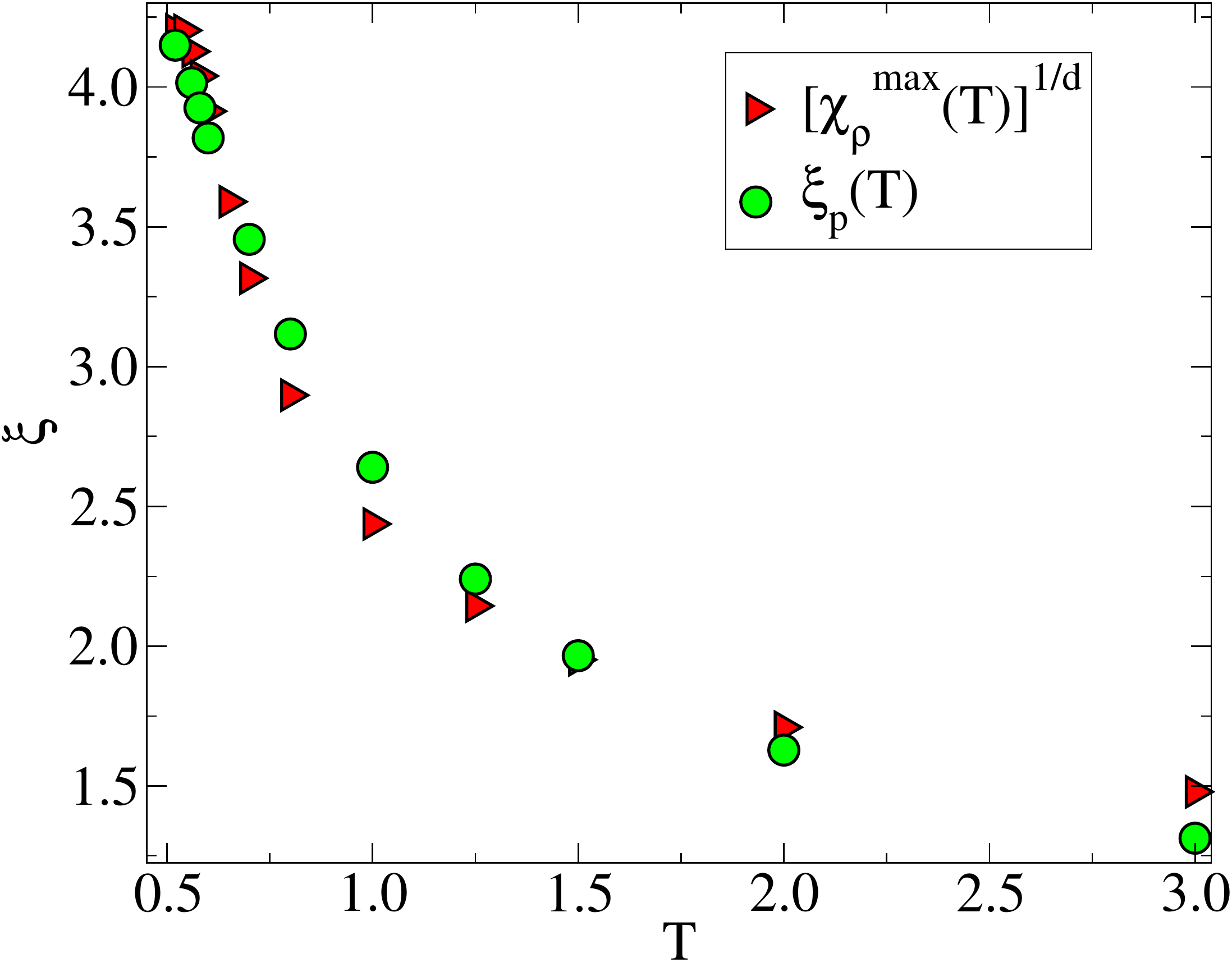}
\hskip -0.1cm
\includegraphics[width=0.24\textwidth]{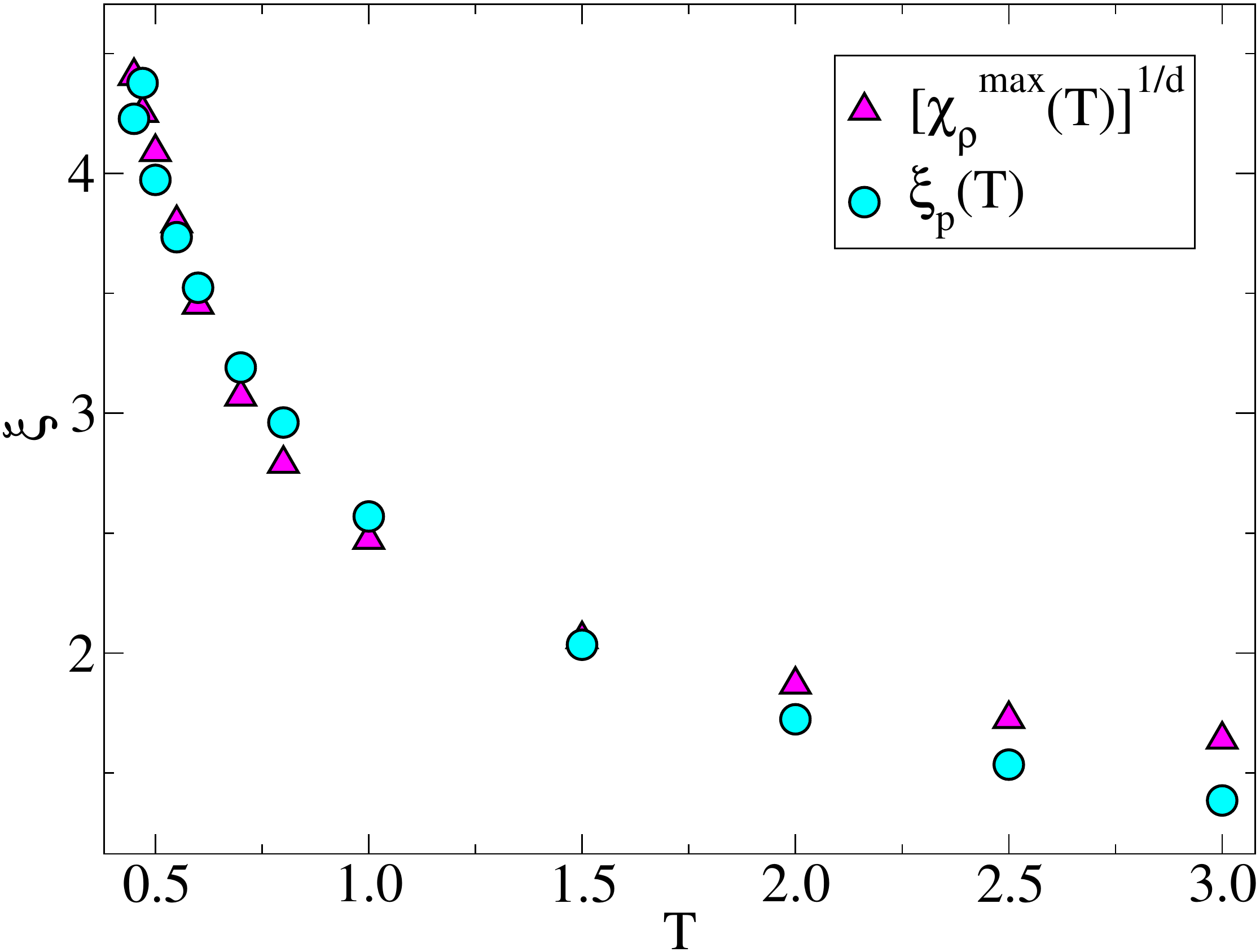}
\vskip 0.1cm
\includegraphics[width=0.235\textwidth]{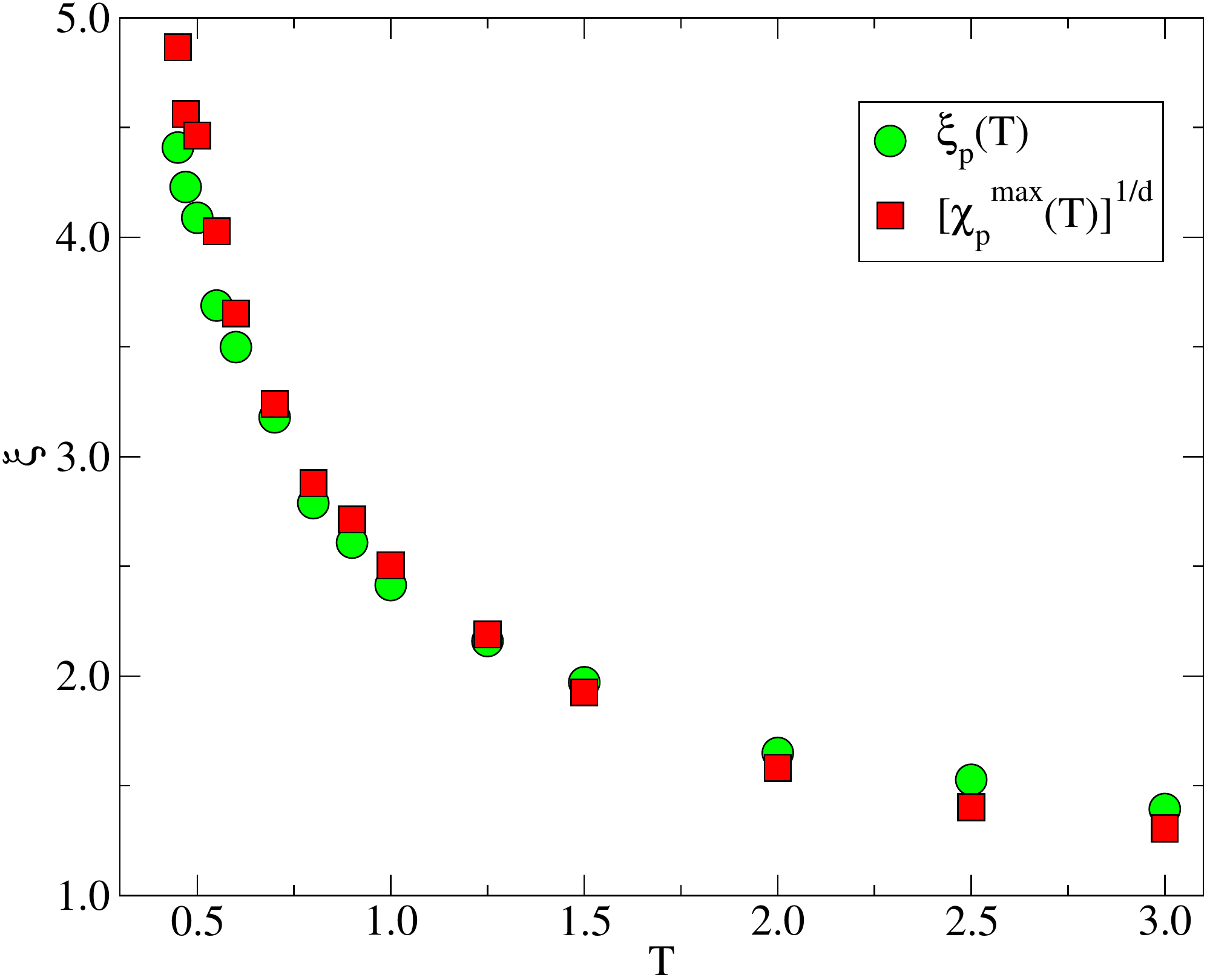}
\includegraphics[width=0.235\textwidth]{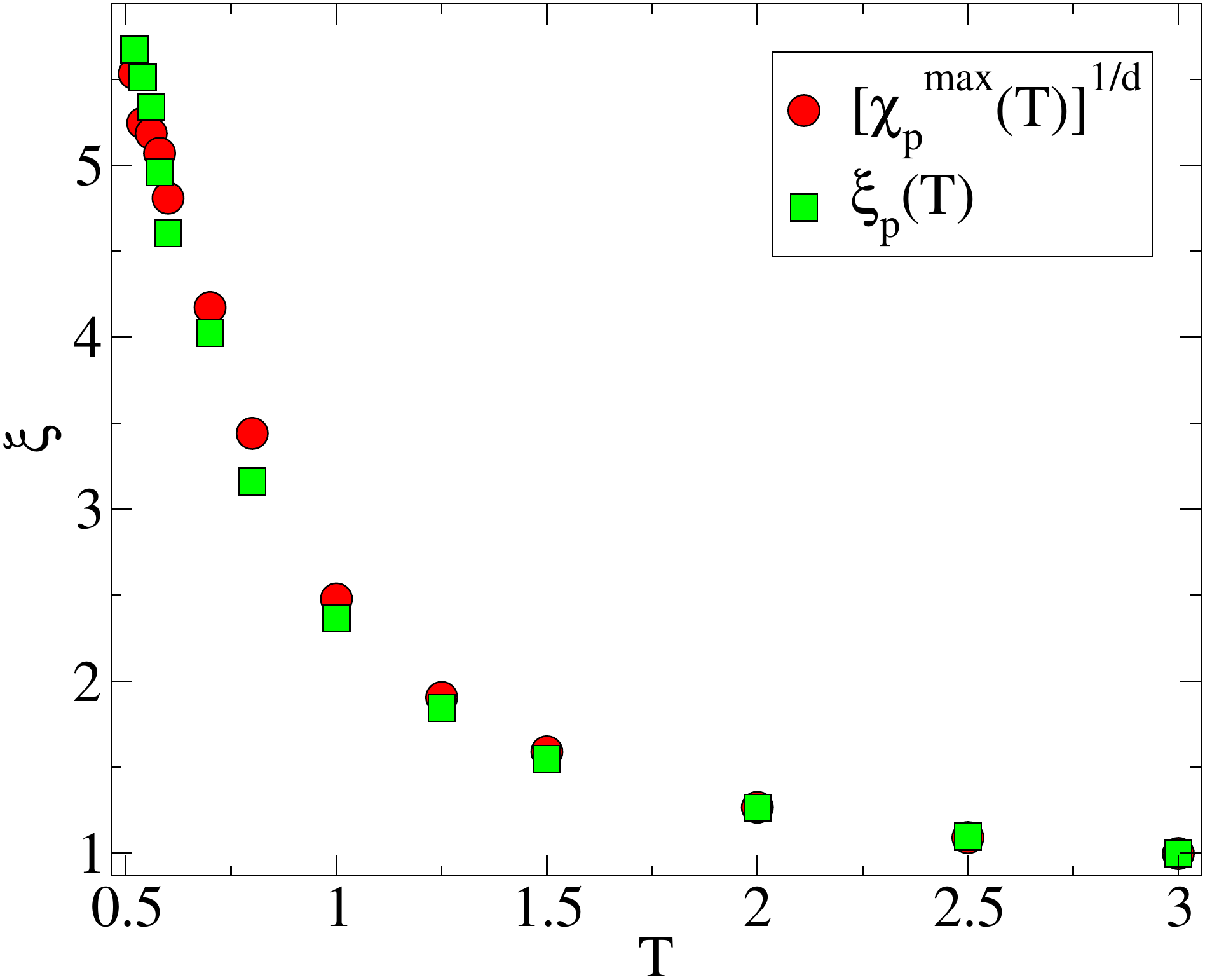}
\caption{{\bf Random Pinning Length Scale:} Comparison of 
$[\chi_p^{max}(T)]^{1/d}$ 
with the pining length-scale $\xi_p(T)$ 
for all the model system studied: 3dKA (top left), 3dR10 (top 
right), 2dmKA (bottom left) and 2dR10 (bottom right).}
\label{chiRhoLengths}
\end{center}
\end{figure}
\begin{figure*}
\begin{center}
\includegraphics[width=0.3\textwidth]{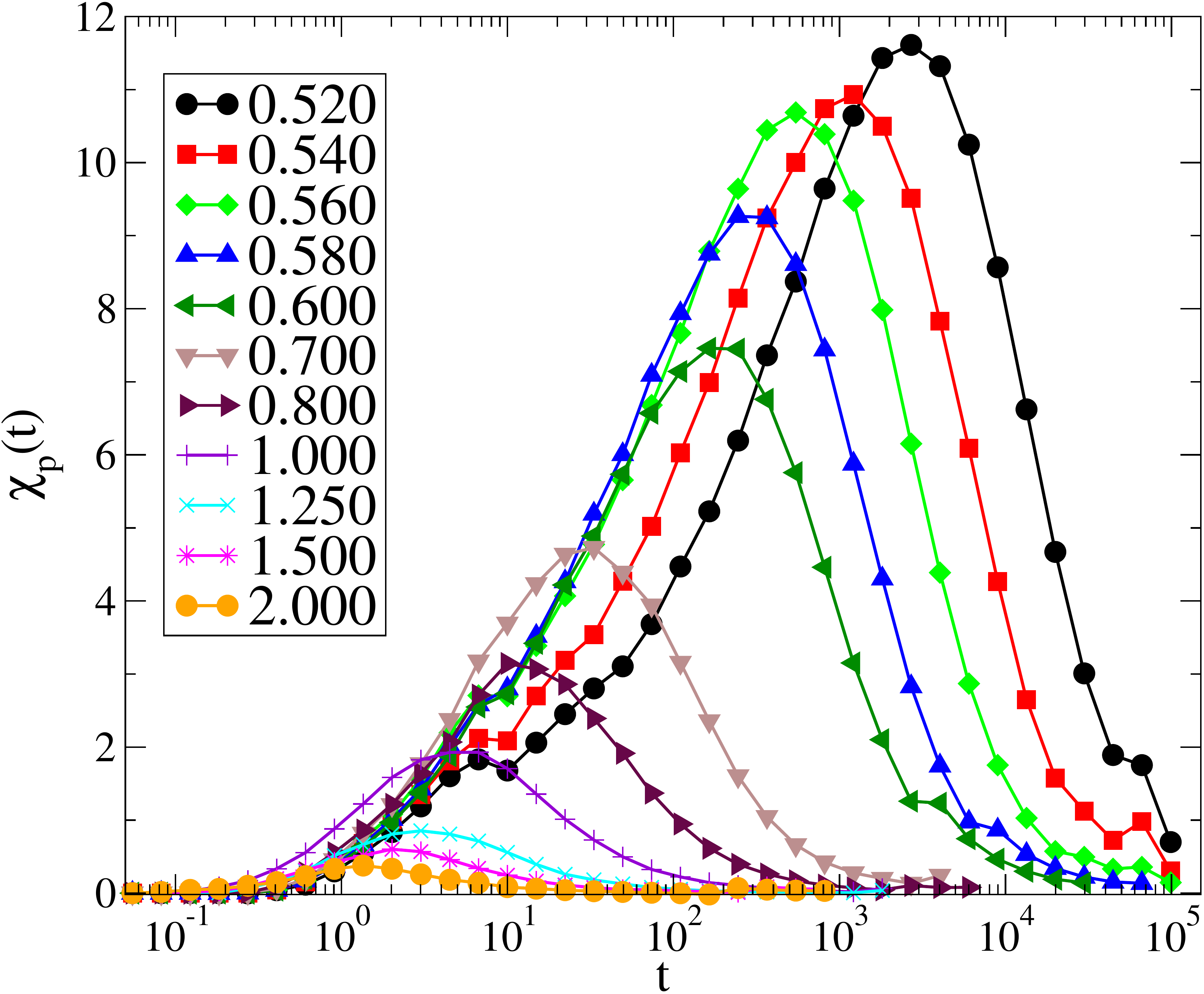}
\includegraphics[width=0.32\textwidth]{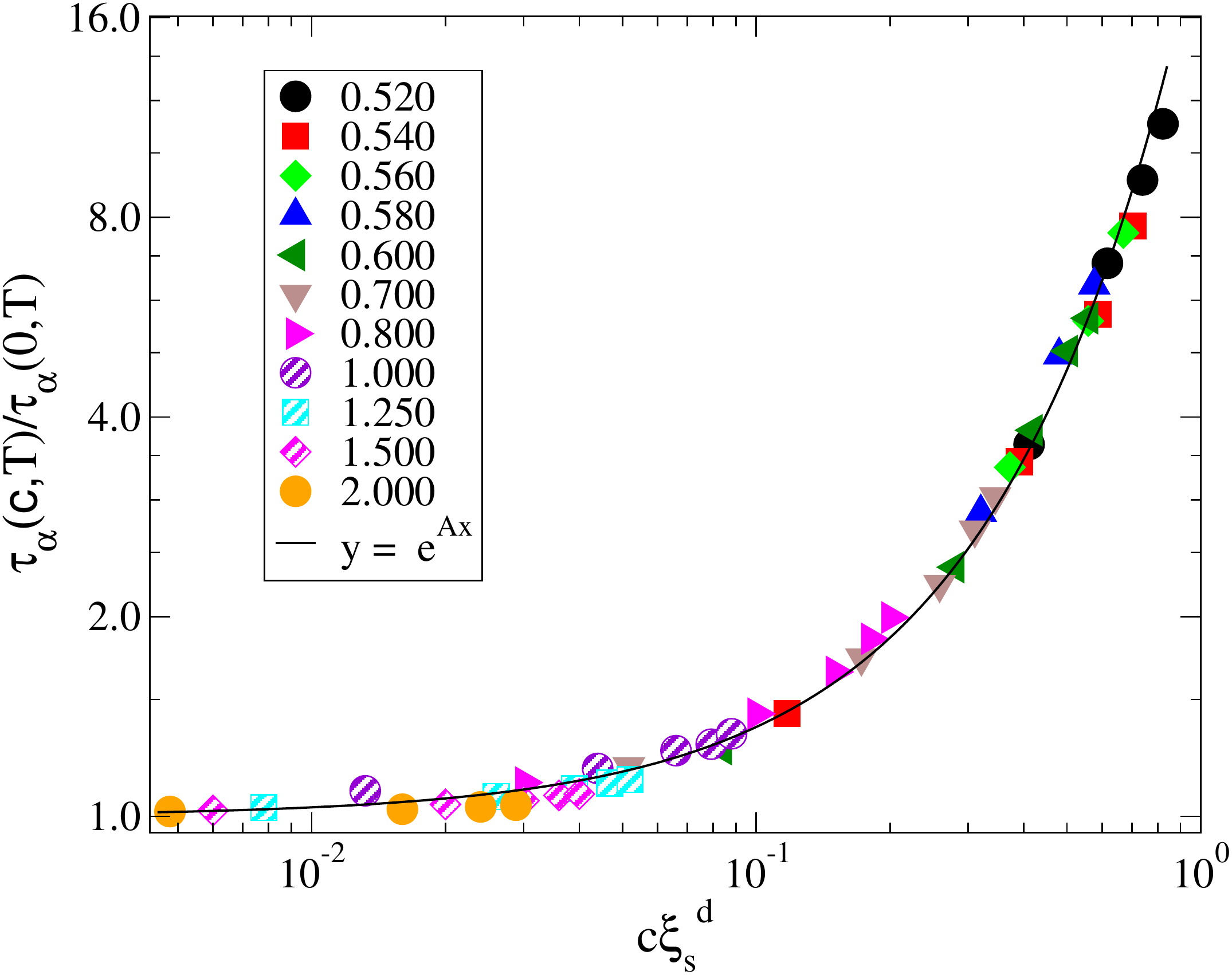}
\includegraphics[width=0.31\textwidth]{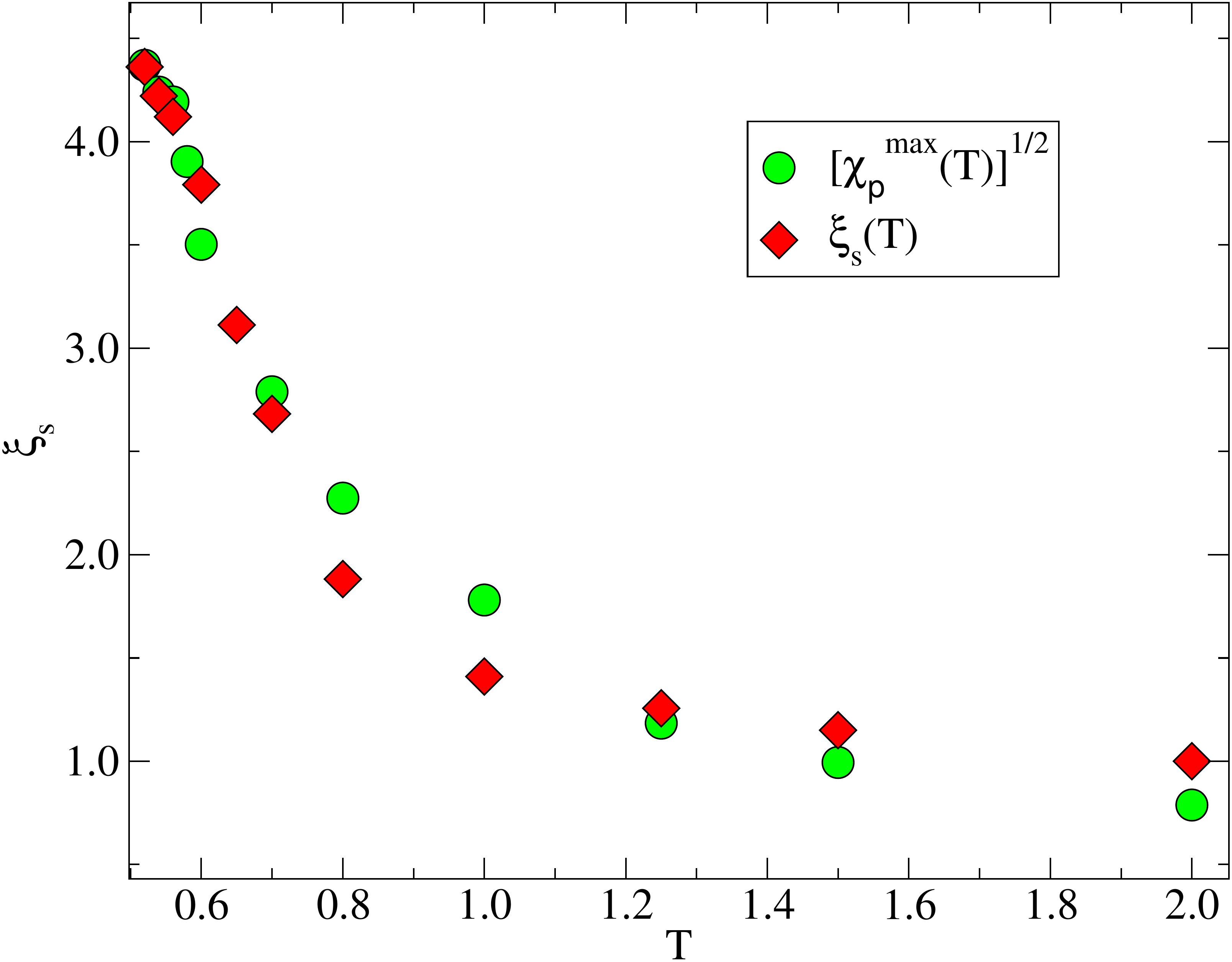}
\vskip 0.2cm
\includegraphics[width=0.32\textwidth]{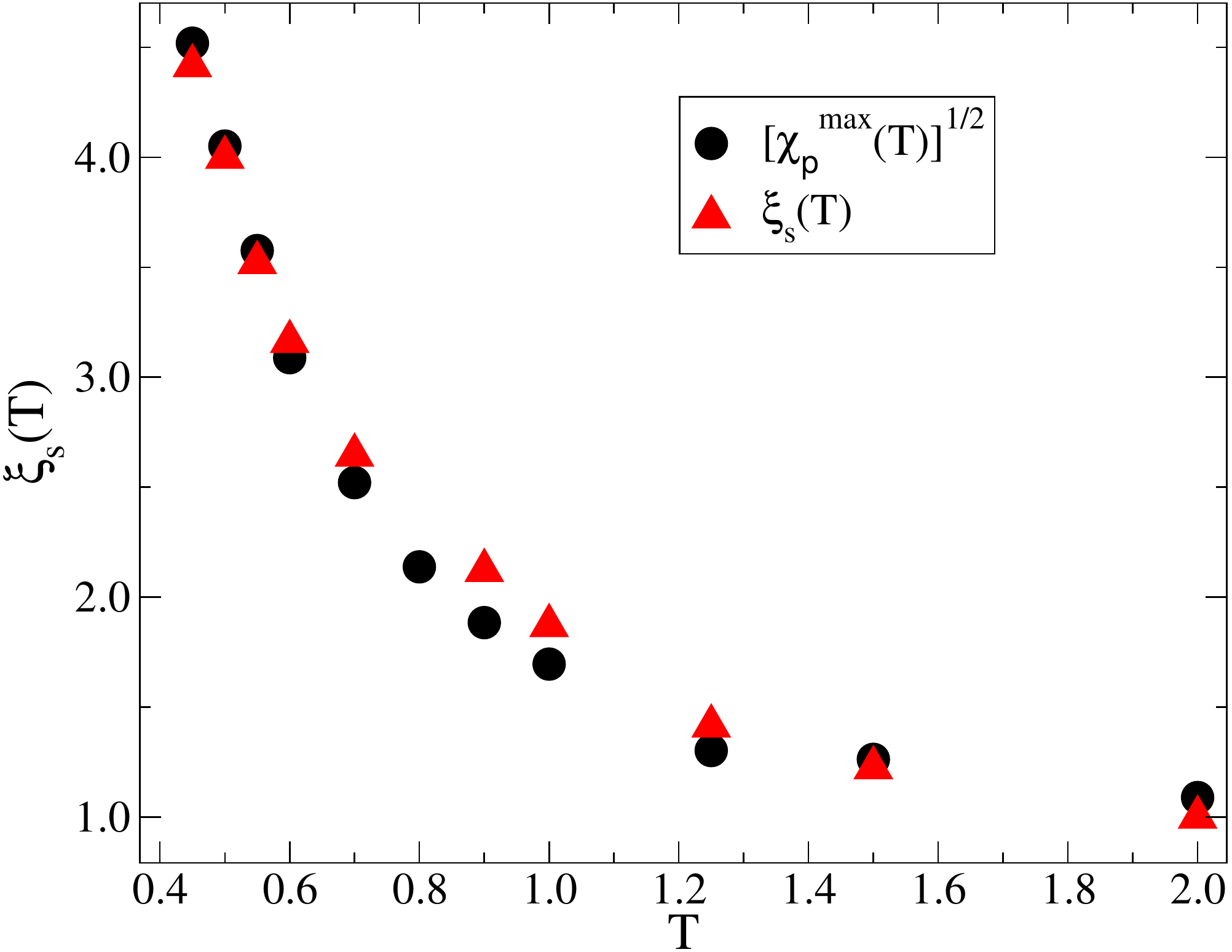}
\includegraphics[width=0.31\textwidth]{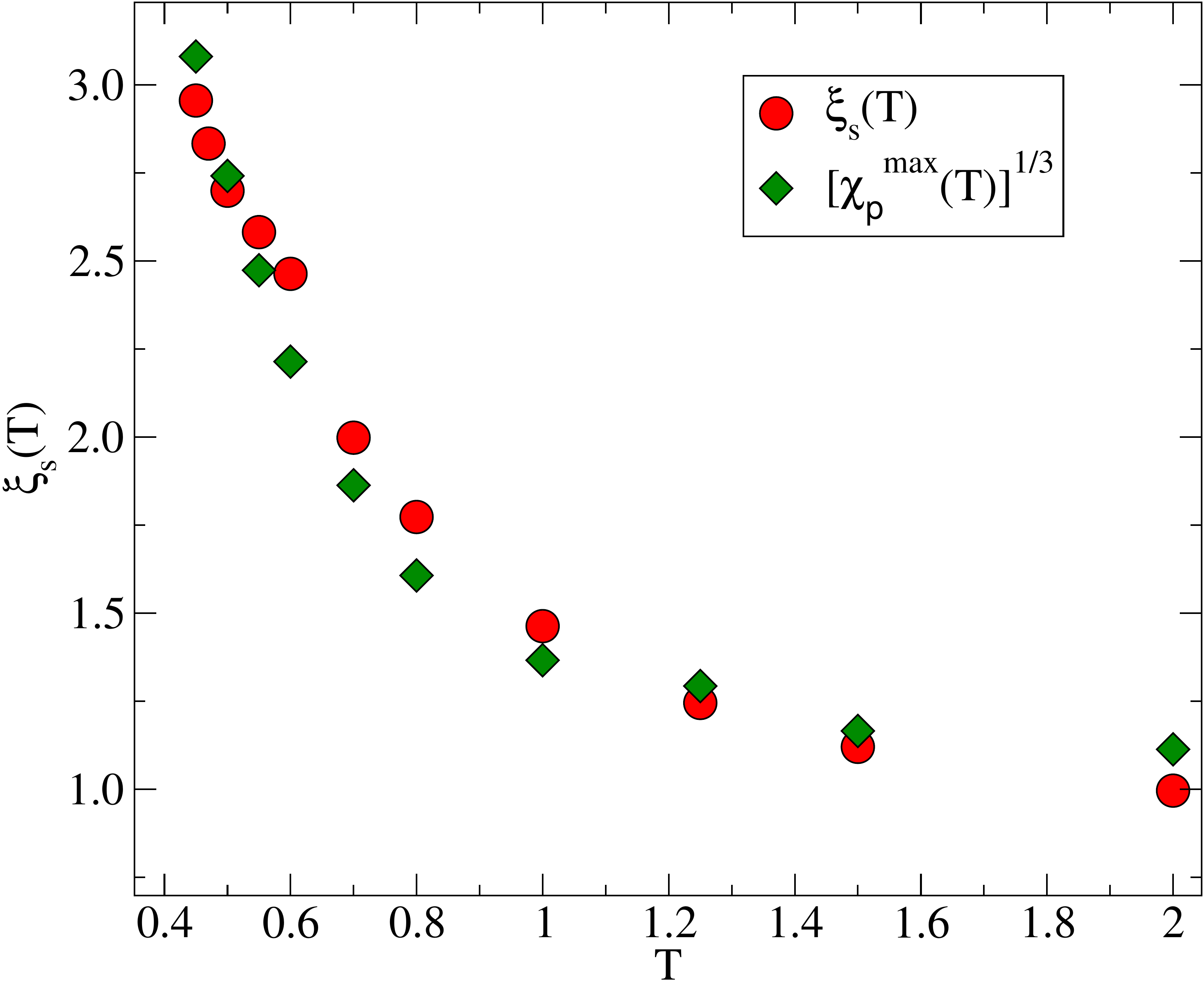}
\includegraphics[width=0.31\textwidth]{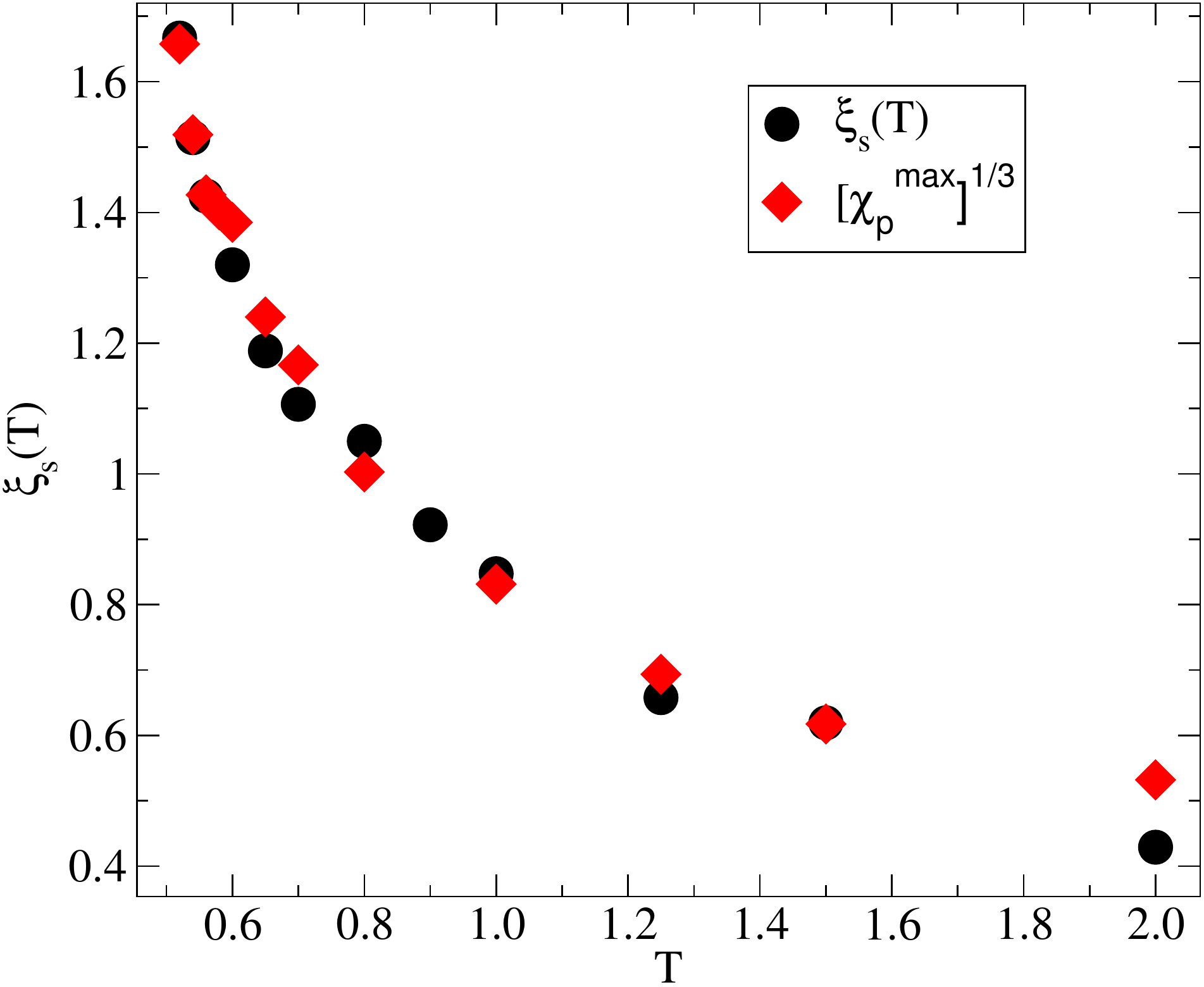}
\caption{{\bf Length-scale of Static Amorphous Order:} (top left) The 
variation of pinning susceptibility, $\chi_p(t)$ with time for various 
temperatures for the 2dR10 model using heavy ``solute'' particles 
instead of strictly pinned particles. (top middle) Validation of the proposed 
scaling ansatz. 
Comparison of the peak height of pinning
susceptibility with the static length scale of amorphous order for different
models studied: 2dR10 (top right), 2dmKA (bottom left), 3dKA (bottom middle)
and 3dR10 (bottom right). Note that the static length-scale, $\xi_s$ is being
calculated independently using multiple conventional methods (see text for
details). Very good agreement of data with conventional method clearly 
establishes the generic nature of the proposed method for studying 
growth of amorphous order in supercooled liquids approaching glass transition.}
\label{lengthChiRhoMD3dKAR10}
\end{center}
\end{figure*}
Thus, the peak value of the $\chi_p(t)$ gives 
us a direct measure of the pinning length-scale $\xi_p$. We have checked 
that the results do not depend on the specific value of $c$ chosen in 
the calculation as long as it is small (see Ref. \cite{SI} for further discussion).
In both the panels of Fig.~\ref{chiRhoLengths}, we have plotted 
$[\chi_p^{max}]^{1/d}$ and $\xi_p$ against the temperature for
the four models studied. It is clear that $[\chi_p^{max}]^{1/d}$ is
indeed proportional to the static length scale $\xi_p$.

Motivated by this observation, we next tried to see whether the concept 
of random pinning applies to a broader context. Instead of actually 
pinning a set of randomly chosen particles we have introduced a small number 
of slightly larger sized particles (solute). If the diffusion
constants of these solute particles are much smaller than those of the
solvent particles, then they should produce effects similar to pinned particles.
Indeed it turns out that small concentration of larger sized solute particles 
do produce dynamical effects similar to random pinning. This is
extremely useful since the implementation of random pinning potentials
in experiments in three dimensions can be quite challenging if not
impossible. 

In the top left panel of Fig. \ref{lengthChiRhoMD3dKAR10}, we plot
such a pinning susceptibility for the 2dR10 model. This susceptibility
has very similar behavior 
as that of the random pinning and similar scaling arguments can be given to 
rationalize the observations with the static length scale, $\xi_s$ being the 
dominant length scale. In top middle panel of Fig.~\ref{lengthChiRhoMD3dKAR10}, 
we have shown such a scaling collapse for the relaxation time. In different 
panels of Fig.~\ref{lengthChiRhoMD3dKAR10}, we have compared the temperature 
dependence of the peak height of $\chi_p(t)$ for 2dR10, 2dmKA, 3dKA and 3dR10 
model with $2\%$ of larger sized solute particles added to the original glass forming 
liquids, with that of the static length scale obtained using combination of 
three methods : {\bf (a)} Point-to-Set method, {\bf (b)} Finite size scaling 
of relaxation time and {\bf (c)} Finite size scaling of minimum eigenvalue 
(see Ref. \cite{SI} for further details about these methods and related 
results). Notice that the static length scale compared here is not the pinning 
length scale rather it is the static length scale of amorphous order, $\xi_s$. 
This is because $\xi_s$ is the only static length scale which governs the 
relaxation barriers in the system without any random pinning. The 
comparison between peak height of pinning susceptibility, $\chi_p^{max}$  
and the static length scale, $\xi_s$ in Fig.~\ref{lengthChiRhoMD3dKAR10} for 
all the models are indeed very good. 

We would like to emphasize that when one introduces larger sized
solute particles in the system, the effective packing fraction will increase
and in effect the pressure will also increase. To remove the effect arising 
from the change in effective packing fraction,
we have maintained the same pressure as that 
of the original binary mixture at each temperature in our simulations
(see Ref. \cite{SI} for further discussion). In Ref. \cite{SI}, we have presented 
mathematical arguments for existence of the scaling form, Eq.~\ref{scalingFn}
and it is important to mention that Pinning Susceptibility indeed obeys the
mathematically required constraints to have the same scaling form. On the 
other hand, $\chi_{\phi}$ or $\chi_T$ may show similar increase of peak 
height with increasing density or decreasing temperature, but they do not 
satisfy the required conditions for the existence of similar scaling function
(see Ref. \cite{SI} for an in-depth discussion on this issue). 
We therefore
believe that 
conventional susceptibilities are not equivalent to our proposed pinning
susceptibility. Although 
the scaling  arguments we
  have proposed
 rationalize our 
observations of a possible connection between the
pinning susceptibility and 
the static length scale of amorphous order, a detailed microscopic theory 
is essential to understand the the observed results.

The Inhomogeneous Mode Coupling Theory (IMCT) \cite{06BBMR}, an extension of 
the standard Mode Coupling Theory (MCT) \cite{mct1,mct} to partially include 
relevant fluctuations, suggests application of a 
spatially localized field which couples to density fluctuations to estimate 
the
increase in cooperativity
in a supercooled liquid. This theory predicts a power law growth of
the correlation volume while approaching the glass transition, the
correlation volume being dynamic in nature.  Similar extension of MCT
for liquids with small concentration of solute particles whose
diffusion coefficients are much smaller than the solvent particles may
reveal the observed microscopic connection between the pinning
susceptibility and amorphous order.

In conclusion, we have proposed a new susceptibility, the pinning 
susceptibility, which can directly measure the length scale
associated with 
amorphous order in supercooled liquids. The 
main virtue of this method is its simplicity and robust applicability to a 
wide variety of systems. We believe that this is the first proposal of 
the much sought after experimentally realizable correlation function which 
directly picks up the size of the correlated region in glass forming liquids. 
All previous propositions involve estimating the correlation length of 
amorphous order in an indirect way. This susceptibility can be measured in 
molecular glasses without much difficulty by introducing a small 
concentration of solute molecules which are somewhat larger than the the 
solvent molecules. Extensions to colloidal glass formers will also be fairly 
simple. This new method carries the potential to establish the 
possible connection between the growth of amorphous order with the rapid 
increase of viscosity in experimentally relevant glass forming liquids 
and may help us understand the puzzles of glass formation and the glass 
transition in near future.
  
\begin{acknowledgments}
We would like to thank Chandan Dasgupta, Giulio Biroli, Srikanth Sastry, Sriram 
Ramaswamy and Surajit Sengupta for many useful discussions. 
\end{acknowledgments}

\clearpage\pagebreak


\section*{Supplementary material}

# Pinning Susceptibility : A Novel Method to Study Growth of Amorphous Order in Glass-forming Liquids – Supplementary Information

Rajsekhar Das[1], Saurish Chakrabarty[2], and Smarajit Karmakar[1]

[1] Centre for Interdisciplinary Sciences, Tata Institute of Fundamental Research,
21 Brundavan Colony, Narisingi, Hyderabad, 500075, India,

[2] International Centre for Theoretical Sciences, Tata Institute of Fundamental Research,
Shivakote, Hesaraghatta, Hubli, Bangalore, 560089, India.

This supplementary information is arranged as follows. In Sec. , we introduce the systems we studied and the simulation details. In Sec. , we give some standard definitions of quantities referred to in the main text related to the physics of dynamical heterogeneity (DH) in supercooled liquids. In Sec. , we describe the methods we used to obtain the static length-scale $\xi_s$. In Sec. we discuss details about the scaling arguments presented in main article. In Sec. , we present a new argument, based on our obtained values of the pinning susceptibility, about the possibility of an ideal glass state at some critical value of the pinning concentration. We also outline a new explanation of the scaling form of $\tau_\alpha(c, T)$ on pinning concentration $c$ used in Refs. [1] and [2]. The arguments are based on the non-singular dependence of the peak height of the pinning susceptibility on the pinning concentration. Finally in Sec. , we discuss the dependence of $\chi_p(t)$ on the choice of $\delta c$.

## MODELS AND THE SIMULATION DETAILS

We study the following four model glass-forming liquids:

**3dKA** – The Kob-Andersen mixture [3]. A $80 : 20$ binary mixture of particles interacting via Lennard-Jones potentials in three spatial dimensions.

$$V_{\alpha\beta}(r) = 4\epsilon_{\alpha\beta} \left[ \left( \frac{\sigma_{\alpha\beta}}{r} \right)^{12} - \left( \frac{\sigma_{\alpha\beta}}{r} \right)^6 \right]. \qquad (1)$$

Here, $\alpha, \beta \in \{A, B\}$ and $\epsilon_{AA} = 1.0$, $\epsilon_{AB} = 1.5$, $\epsilon_{BB} = 0.5$, $\sigma_{AA} = 1.0$, $\sigma_{AB} = 0.80$, $\sigma_{BB} = 0.88$. The number density of particles is 1.2 and the temperature range is $T \in \{0.45, 3.0\}$. The interaction potential is cut off at $2.50\sigma_{\alpha\beta}$ using a quadratic polynomial such that the potential and its first two derivatives are continuous at the cutoff distance.

**2dMKA** – The same system in two dimensions with the two types of particles having proportion $65 : 35$ instead of $80 : 20$.

**3dR10** – A $50 : 50$ binary mixture [11] interacting in three dimensions via the potential,

$$V_{\alpha\beta}(r) = \epsilon_{\alpha\beta} \left( \frac{\sigma_{\alpha\beta}}{r} \right)^{10}. \qquad (2)$$

Here, $\epsilon_{\alpha\beta} = 1.0$, $\sigma_{AA} = 1.0$, $\sigma_{AB} = 1.22$ and $\sigma_{BB} = 1.40$. The potential is cut off at $1.38\sigma_{\alpha\beta}$ using a quadratic polynomial to make the potential and its first two derivatives continuous at the cutoff. The number density of particles is 0.81 and the temperature range is $T \in \{0.52, 3.0\}$.

**2dR10** – The same system as above in two dimensions but with a number density of 0.85.

We have performed NVT molecular dynamics simulations for all the model systems where modified leap-frog algorithm was used for the integration and Berendsen thermostat to keep the temperature constant.

**Random Pinning Protocol:** A fraction of particles are chosen randomly from an equilibrated configuration of the the studied model system and then their dynamics are frozen. To study the dynamics of the system at varying pinning concentration we have changed the concentration in the range $c \in \{0.005, 0.200\}$ [1, 2]. Subsequently scaling analysis was done to obtain the random pinning length scale, $\xi_p$ for all the model system studied. $\xi_p$ data presented in this work are taken from Refs. [1, 2]. Please refer to Refs. [1, 2] for complete details on the dynamics of supercooled liquids with random pinning.

**Ternary Mixture:** In order to generate dynamical effects which are similar to random pinning, we have introduced a small fraction of a third type of particle with a bigger diameter for all the four systems studied. We will refer to these systems as ternary version of the respective models. Interaction potential for 3dKA and 2dMKA systems is same as given in Eqn.1, but here, $\alpha, \beta \in \{A, B, C\}$ and $\epsilon_{AA} = 1.0$, $\epsilon_{AB} = 1.5$, $\epsilon_{AC} = 1.0$, $\epsilon_{BB} = 0.5$, $\epsilon_{BC} = 1.0$, $\epsilon_{CC} = 1.0$, $\sigma_{AA} = 1.0$, $\sigma_{AB} = 0.80$, $\sigma_{AC} = 1.10$, $\sigma_{BB} = 0.88$, $\sigma_{BC} = 1.04$, $\sigma_{CC} = 1.20$. For 3dR10 and 2dR10 systems the interaction potential is given by Eqn. 2, and $\epsilon_{\alpha\beta} = 1.0$, $\sigma_{AA} = 1.0$, $\sigma_{AB} = 1.18$, $\sigma_{AC} = 1.50$, $\sigma_{BB} = 1.40$, $\sigma_{BC} = 1.70$, $\sigma_{CC} = 2.0$. Fraction of the particle chosen as third type of particles is $c \in \{0.005, 0.080\}$ for all the four systems. NPT molecular dynamics simulations are performed for the ternary models at the pressures obtained from the corresponding binary model simulations for respective temperatures.

## DYNAMIC CORRELATION FUNCTIONS

The dynamic susceptibility is a measure of the fluctuations in the two point overlap correlation function defined

as

$$Q(T, c, t) = \frac{1}{(1-c)N} \left[ \left\langle \sum_{i=1}^{(1-c)N} w\left(|\vec{r}_i(t) - \vec{r}_i(0)|\right) \right\rangle \right]. \tag{3}$$

The summation is over the $(1-c)N$ unpinned particles. The window function $w(x) = \Theta(a - x)$ where $\Theta(x)$ is the Heaviside step function. The window function is chosen to remove from the correlation function, the effects of short time vibrational motion of a given particle inside the "cage" formed by its neighboring particles. The value of the parameter $a$ is chosen to be 0.30 for this study but qualitative results are not affected by any particular choice of this parameter [2, 4]. The square brackets indicate averaging over different realizations of the pinned particles and angular brackets stand for ensemble average.

Here we will denote overlap correlation function for the system with no random pinning simply as $Q_0(t) \equiv Q(T, c = 0, t)$. Note that explicit temperature dependence $T$ is not mentioned for simplicity. With these definition the four-point dynamic susceptibility will be given as

$$\chi_4(t) = \frac{1}{N} \left[ \langle Q_0^2(t) \rangle - \langle Q_0(t) \rangle^2 \right] \tag{4}$$

The four-point structure factor is given by,

$$S_4(k, t) = \frac{1}{N} \langle |\sum_i e^{i\vec{k} \cdot \vec{r}_i(0)} w(|\vec{r}_i(t) - \vec{r}_i(0)|)|^2 \rangle \tag{5}$$

Please see Ref. [8] for a comprehensive review on the dynamical behavior of $\chi_4(t)$ and $S_4(q, t)$ and their relations to dynamic heterogeneity length scale, $\xi_d$.

## METHODS OF CALCULATING THE STATIC LENGTH-SCALE, $\xi_s$

Although there is a consensus on the existence of a static length-scale in supercooled liquids that grows as the temperature is lowered, obtaining it may be quite challenging. In this work, we have used the following three methods to obtain the static length-scale.

### Point-To-Set Method:

The point-to-set (PTS) length-scale was introduced by Bouchaud and Biroli in Ref. [5] and it was numerically calculated for the first time in Ref. [7] in a model binary glass former. This method is useful at somewhat high temperatures when the length-scale is not too large. The main difficulty comes from the dynamical procedure involved in calculating this length scale. In the PTS method, one needs to take an equilibrated configuration

at a given temperature and then define a smaller spherical region of radius, $R$. The particles outside this spherical cavity are pinned at their respective positions and only the particles inside the cavity are allowed to evolve according to Newtonian dynamics. The static overlap correlation is calculated for the particles inside the cavity and in order to remove boundary effects the static overlap is calculated only for those particles which are in the center of the cavity. This procedure is repeated for many different sizes of the cavity and the size dependence of the static overlap is obtained to finally extract the point-to-set length scale.

The external pinned boundary makes the dynamics of the internal particles very slow especially for small cavities and one needs to employ a sophisticated sampling method (Swap Monte Carlo Method [14]) to partly overcome this difficulty. As this enhanced sampling method is not generically applicable to any model system, the sampling problems are not always easy to overcome [10] especially at lower temperature. In this study we have used PTS methods only at those temperature regime where it can be done within accessible computation time. Some of the PTS length-scale reported in this work are taken from Ref. [2].

### Finite Size Scaling of Minimum Eigenvalue:

The second method is the finite-size-scaling of the minimum eigenvalue of the Hessian matrix obtained for the inherent structure of the supercooled liquid at a given temperature [11]. This method is expected to work in the low temperature regime where the supercooled liquid spends a considerable amount of time around a given minimum of its free energy landscape and a harmonic approximation works for a finite (but short) time. We will briefly describe the method here. Please refer to Ref. [11] for the details of this method.

Karmakar et al. in [11] argued that the minimum eigenvalue $\lambda_{min}(T)$ of the Hessian matrix obtained at the inherent structure closest to the configuration of a supercooled liquid at given temperature $T$ takes the scaling form,

$$\frac{\langle \lambda_{min}(T) \rangle}{\langle \lambda_D(T) \rangle} = \mathcal{F} \left[ \xi_s^d(T) \left( \frac{1}{N} - \frac{Ad}{2} \left( \frac{\langle \lambda_{min}(T) \rangle}{\langle \lambda_D(T) \rangle} \right)^{d/2} \right) \right]. \tag{6}$$

$\lambda_D(T)$ is the square of the Debye frequency, $d$ is the dimensionality of space, $A$ is an adjustable constant chosen to get the best scaling collapse and $\mathcal{F}$ is an unknown scaling function. In Ref. [11], the validity of this method was demonstrated for a few model glass forming liquids and later in Ref. [12] it was shown that the temperature dependence of the PTS and minimum eigenvalue length-scales is same for various generic glass forming liquids. Recently in Ref. [13], this method was used to

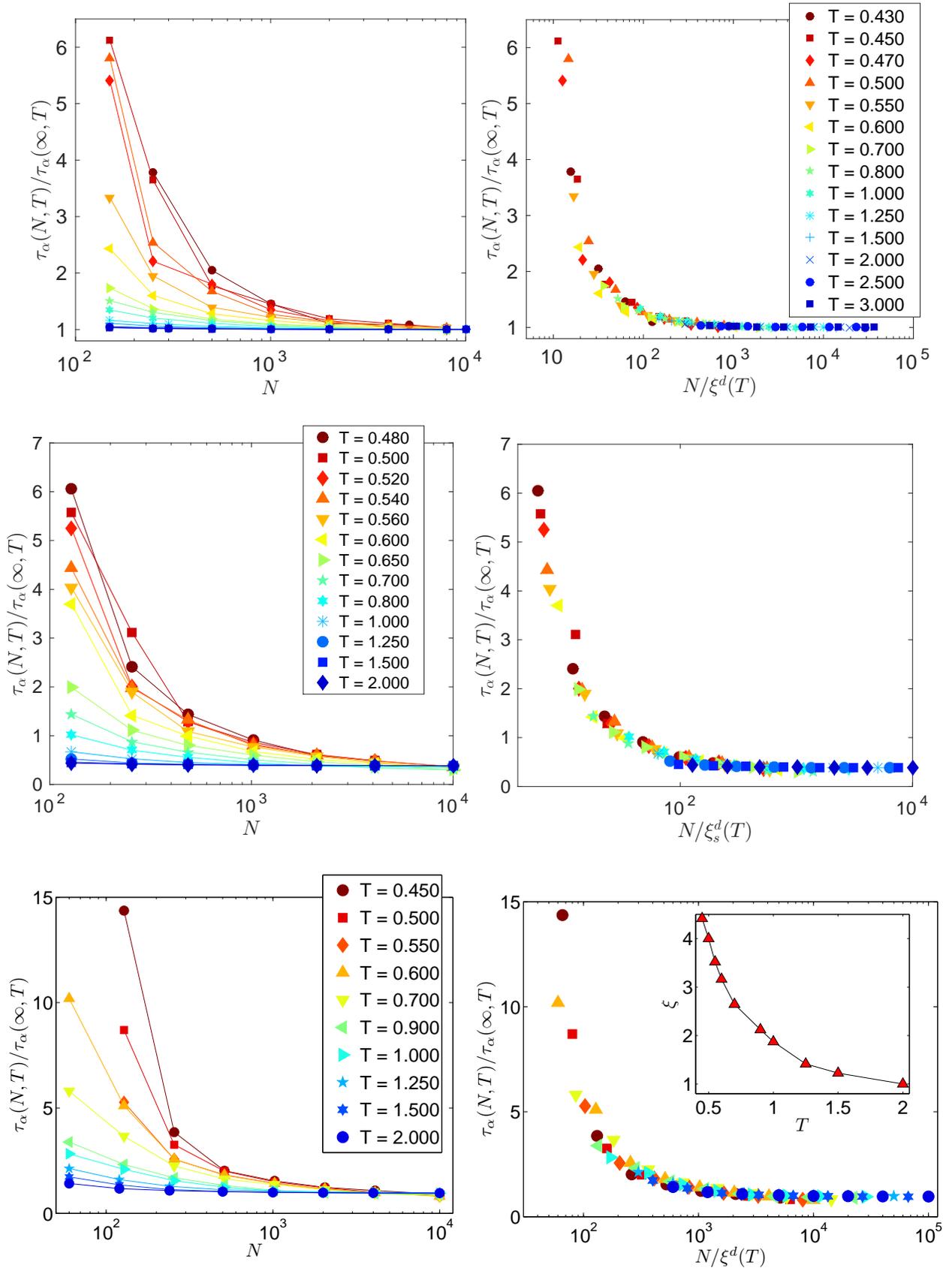

FIG. 1: Top Panel: Finite size effects of $\tau_\alpha$ for the 3dKA model and the corresponding data collapse to obtain static length scale $\xi_s$ shown in the next panel. Similar analysis done for 2dR10 model (middle panel) and 2dmKA model (bottom panel) are also shown. In all these case scaling collapse observed to be very good.

obtain the static length scale for a glass forming liquids for which equilibrium state can be relatively easily sampled using the Swap Monte Carlo technique. A large change in length-scale was reported in that work.

**Finite Size Scaling of $\alpha$-relaxation time**

The third method we use is the finite-size-scaling of the $\alpha$-relaxation time, $\tau_\alpha$. The $\alpha$-relaxation time of supercooled liquids shows a strong system size dependence. It decreases monotonically with increasing system size before attaining its asymptotic value at the thermodynamic limit. The dependence is however controlled by a single static length-scale and the following scaling form is used to obtain it.

$$\frac{\tau_\alpha(N,T)}{\tau_\alpha(\infty,T)} = \mathcal{G}\left(\frac{N}{\xi_s^d(T)}\right). \qquad (7)$$

$\mathcal{G}(x)$ is an unknown scaling function. The infinite system size relaxation time $\tau_\alpha(\infty,T)$ and the static length-scale $\xi_s(T)$ are chosen so as to get a good data collapse when $\tau_\alpha(N,T)/\tau_\alpha(\infty,T)$ is plotted against $N/\xi_s^d(T)$ for all temperatures on the same graph. In Fig. 1, we have shown such data collapse to obtain the static length scale for some of the studied systems. In Ref. [6], it is demonstrated that the length-scale obtained from the FSS analysis of $\tau_\alpha$ is indeed the same as the length-scale obtained using PTS and FSS of minimum eigenvalue method. For an in-depth discussion on the importance of various length-scales in the dynamics of supercooled liquids see, *e.g.*, Refs. [6] and [8].

**SCALING ARGUMENTS**

In the main article, we have argued that we can actually calculate the static length scale following scaling arguments very similar to the random pinning situations, by introducing a small fraction of solute particles whose diameters are larger than the solvent particles. These particles, as discussed before, will have smaller diffusion coefficients than the diffusion coefficients of the solvent particles. If the difference between these diffusion coefficients is large, then the solute particles behave like pinned particles over the time-scale of structural relaxation of the solvent particles and one would expect scaling arguments similar to random pinning to hold good as well,

$$\log\left[\frac{\tau_\alpha(c,T)}{\tau_\alpha(0,T)}\right] = f\left[c\xi_s^d(T)\right]. \qquad (8)$$

In order to test this scaling ansatz we have done the systematic studies of relaxation time with increasing fraction of solute particles. In Fig. 2 we have shown such a

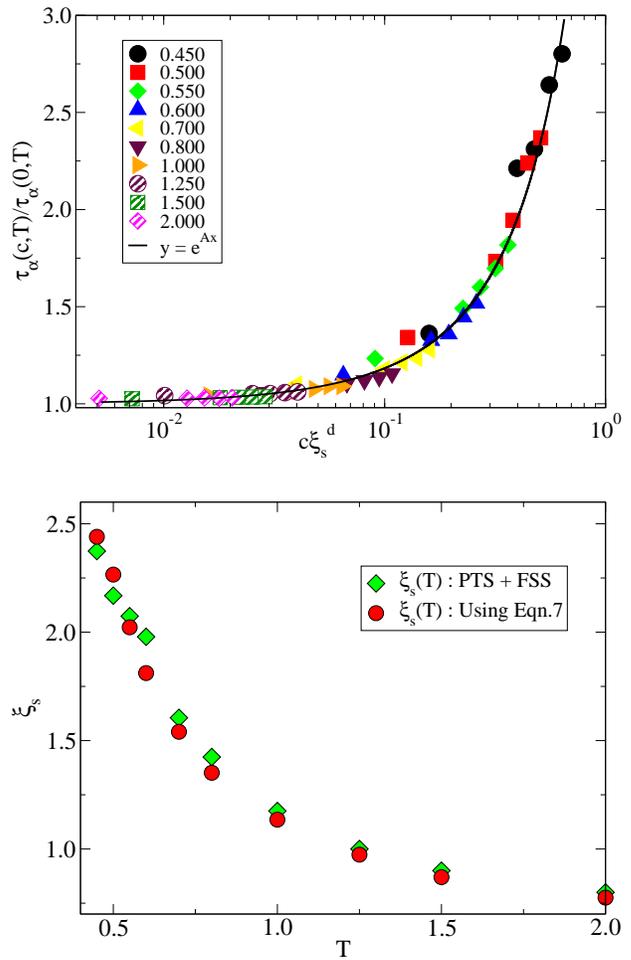

FIG. 2: Top panel: Data collapse using Eq. 8 to obtain static length scale for 3dKA system with third type of particle with bigger diameter. Bottom panle: Comparison of length scales. PTS refer to Point-to-set method and FSS refer to Finite size scaling (see Sec. for details)

scaling analysis for the 3dKA ternary system and compared the obtained length scales with the length scales obtained from finite-size-scaling of the $\alpha$ relaxation time for the same system.

Note that while extracting the functional relationship between the peak height of $\chi_p(t)$ and the static length scale, we have assumed that the peak height of $\chi_p(t)$ appears at a timescale which is proportional to $\alpha$-relaxation time, $\tau_\alpha$. This is an important assumption and without this the scaling arguments will not hold good. To check the validity of this assumptions, we have plotted the time at which $\chi_p(t)$ peaks at different temperatures and $\tau_\alpha$ in the top panel of Fig. 3 for 2dR10 model with random pinning. All other models show exactly the same behaviour. For the ternary model system, again for the scaling arguments to be valid peak of $\chi_p(t)$ should appear at timescale proportional to $\tau_\alpha$. In bottom panel of Fig. 3 we have shown such a comparison for the 2dR10

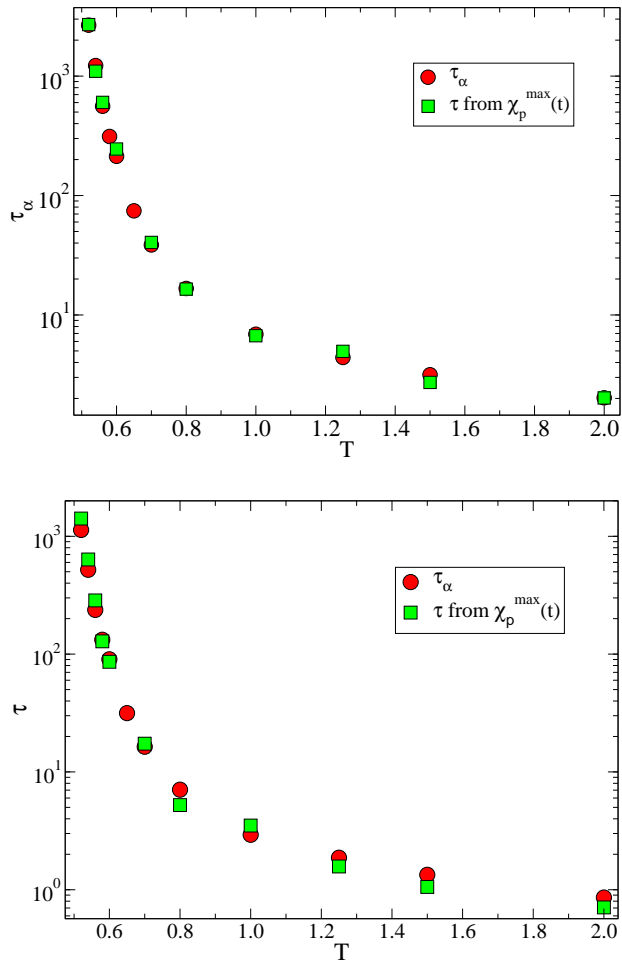

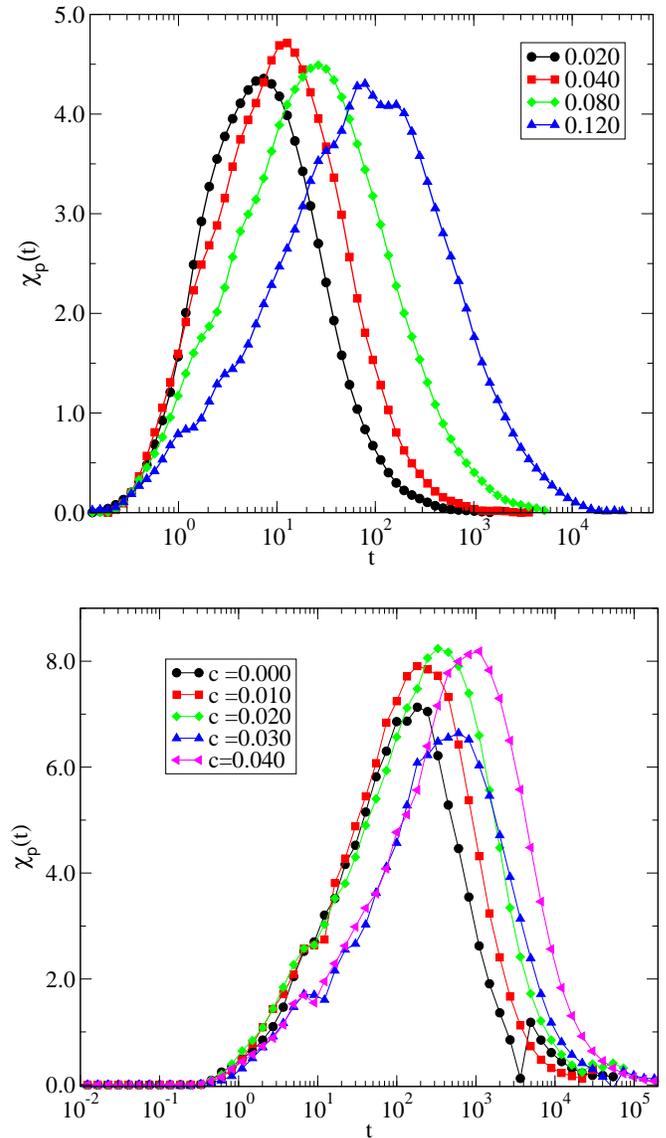

pinning fraction, the overlap correlation function must decay to zero at some finite time. Therefore, there must exist a time-scale $t_0(\delta c)$ such that $Q[c_0(T) - \delta c, T, t] = 0$ for all $t > t_0$. Thus, the pinning susceptibility $\chi_p$ must be infinite for all $t > t_0$. It must be noted that the time-scale at which $\chi_p$ diverges must be shorter than the $\alpha$-relaxation time of the system for any finite $\delta c > 0$ with $t_0(0) \gtrsim \tau_\alpha \to \infty$. From top panel of Fig.4, it is clear that $\chi_p^{max}(T)$ shows no tendency to diverge even when higher and higher pinning fraction data are considered. This gives us a compelling evidence against the possibility of

FIG. 3: The time at witch the maximum of $\chi_p(t)$ appears is compared with $\tau_\alpha$ for 2dR10 model with random pinning (top panel) and for the ternary model (bottom panel).

model. One can clearly see that maximum of $\chi_p(t)$ occurs nearly at $\tau_\alpha$ for all temperatures. Similar results were obtained for other model systems also. These observations clearly validate one of the crucial assumptions made in the scaling theory.

## A NOTE ON THE POSSIBLE IDEAL GLASS TRANSITION WITH RANDOM PINNING

If there is an ideal glass transition as a function of the pinning strength, the pinning susceptibility must diverge for all times larger than some finite time for that value of the pinning strength. This can be proven as follows. Suppose, the ideal glass transition takes place at $c = c_0(T)$ at temperature $T$. In that case, at temperature $T$ and pinning fraction $c_0(T)$, the overlap correlation function must not decay to zero at any finite time. Thus, $Q[c_0(T), T, t] > 0$ for all $t$. However, for a slightly lower

FIG. 4: Top panel: The variation of pinning susceptibility with time for various pinning fractions for the 3dKA model with 2000 particles at $T = 0.80$. Bottom panle: The variation of pinning susceptibility with time for various concentrations of third particles for the ternary version of 2dR10 model with 1000 particles at $T = 0.60$ with $\delta c = 0.01$.

achieving an ideal glass state by just increasing the number of pinned particles in a system. In the bottom panel of Fig. 4, we have shown the variations of peak heights with increasing concentrations of larger third particles for the ternary 2dR10 model to demonstrate the similarity with the pinning susceptibility for random pinning. This clearly shows that peak value of pinning susceptibility does not depend on the pinning concentrations or amount of third particles in the case of ternary models. Similar results are obtained for other model systems also.

## EXISTENCE OF SCALING FUNCTION

In this section, we discuss the criterion for the existence of the observed scaling function. We first find out if there is an optimum value of the pinning fraction, $c_P$, at which the temporal peak of the pinning susceptibility takes its highest value for a given temperature. For this, we set the derivative of $\chi_p[c, T, \tau_\alpha(c, T)]$ with respect to $c$ to zero. Using the stretched exponential form for $Q(c, t)$ (valid at the $\alpha$-relaxation regime), this gives us the following relation:

$$\tau_\alpha \frac{\partial^2 \tau_\alpha}{\partial c_P^2} = \left(\frac{\partial \tau_\alpha}{\partial c_P}\right)^2. \qquad (9)$$

where the temperature dependence of all the quantities are not explicitly written. Solving this, we get,

$$\tau_\alpha(c_P, T) = A(T) \exp\left[B(T)c_P\right]. \qquad (10)$$

A corollary to this is that if the peak height of $\chi_p(t)$ for a given temperature is independent of $c$, then the derivative considered to get the maximum in the above analysis vanishes for all values of $c$. In that case, we get after renaming $A(T)$ and $B(T)$ appropriately,

$$\tau_\alpha(c, T) = \tau_\alpha(0, T) \exp[c\xi_p^d(T)]. \qquad (11)$$

Thus it is essential for the existence of the proposed scaling function that the peak height of $\chi_p(t)$ for a given temperature is independent of the pinning concentration, $c$. In top panel of Fig. 4, we have plotted $\chi_p(t)$ for different pinning concentrations and one observes that the peak height does not grow with increasing pinning concentration. It is rather constant within the numerical statistical error in our data. Thus peak height of $\chi_p(t)$ can be considered to be independent of the pinning concentration. This directly suggests that the temperature at which the ideal glass transition occurs is independent of the pinning fraction as assumed in Ref. [1]. Similarly in the bottom panel of Fig. 4, we have shown similar data for the ternary model of 2dR10. Here also one can clearly see that with increasing concentration of larger size third particles, the peak height of pinning susceptibility does not change. Thus, it may be safe to expect a similar scaling function to hold good for this case also.

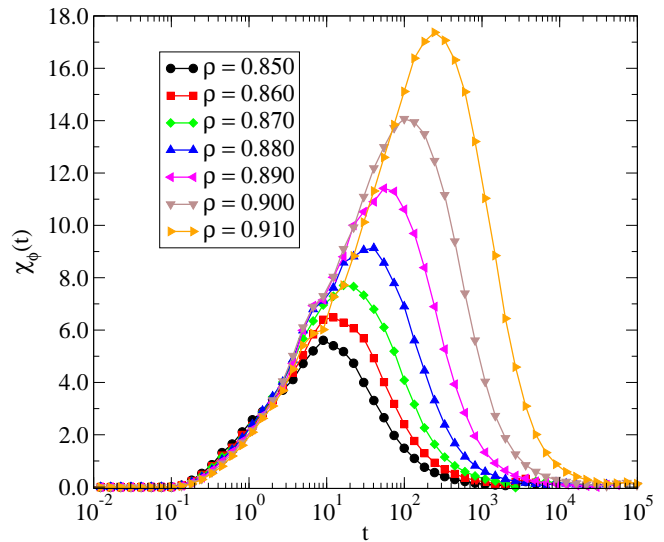

FIG. 5: The variation of $\chi_\phi(t)$ with time for various number density for the 2dR10 model with 1000 particles at $T = 0.80$ with $\delta\rho = 0.01$. The peak height increasing quite strongly with increasing density. This is stark contrast with the pinning susceptibility where peak height does not change with increasing pinning concentrations.

On the other hand, instead of introducing a third type of particle particle (or random pinning) if we simply increase the number density $\rho$ and calculate $\chi_\phi$ at different number density we see that peak height of $\chi_\phi(t)$ increases very strongly with the increase of $\rho$ as shown in Fig. 5 for the 2dR10 model. This is completely different than the pinning susceptibility where the peak height does not change with increasing pinning concentrations. Thus one can rule out similar scaling function to exist for $\chi_\phi$. We would also like to mention that the peak does not appear at a time scale proportional to $\tau_\alpha$ for a given density as a function of temperature. This also rules out the possibility of similar scaling function to exist for $\chi_\phi(t)$.

Thus, we feel that our proposed pinning susceptibility is unique in its own way to directly estimate the growth of amorphous order in supercooled liquids and the above mentioned scaling arguments clearly suggest that the peak height is directly proportional to the correlation volume without any unknown exponents.

## DEPENDENCE OF $\chi_p^{max}(T)$ ON $\delta c$

In this section we have checked the dependence of $\chi_p^{max}$ for various temperature on the choice of $\delta c$ while calculating $\chi_p^{max}$ via numerical differentiation. In Fig. 6, we have plotted $\chi_p^{max}$ using different values $\delta c$ for our model systems. One can clearly see that the peak height does not depend explicitly on the choice of $\delta c$ as long as it is

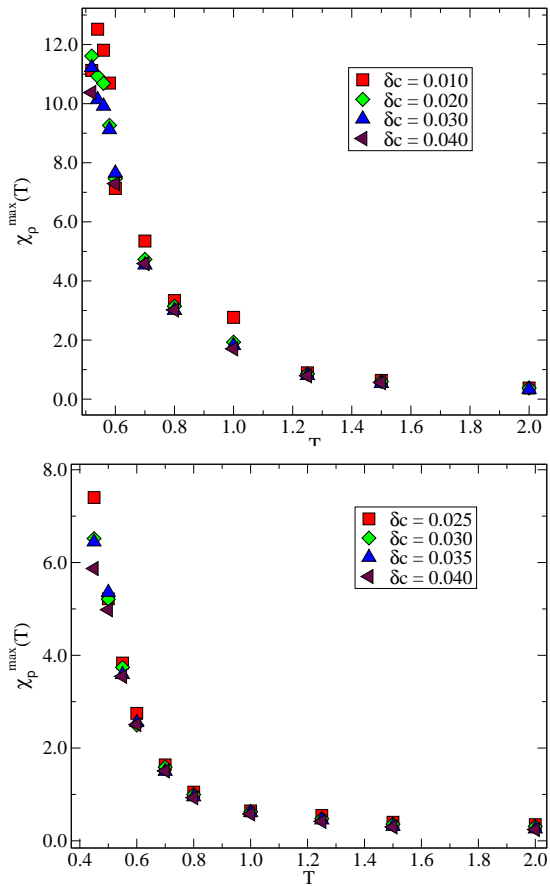

FIG. 6: Top panel: $\chi_p^{max}$ for different choices of $\delta c$ for 2dR10 Ternary model. Notice that up to $\delta c = 0.04$, there is not much dependence of $\chi_p$ on $\delta c$. Bottom Panel: Similar plot for 3dKA Ternary model. One can clearly see that the peak value of the $\chi_p^{max}$ does not depend on the choice of the $\delta c$ for the studied models.

small. This also directly establishes the importance of our proposed susceptibility along with behaviour.



\end{document}